\documentclass[11pt]{article}
\usepackage{amsmath,amssymb,color,epsfig,cite}
\usepackage{mathrsfs}

\usepackage{mathtools}
\usepackage{amsfonts}
\newcommand{\hoch}[1]{$\, ^{#1}$}

\makeatletter
\@addtoreset{equation}{section}
\makeatother

\newcommand{\be}{\begin{equation}}
\newcommand{\ee}{\end{equation}}
\newcommand{\bea} {\begin{eqnarray}}
\newcommand{\eea}{\end{eqnarray}}
\newcommand{\nn}{\nonumber}

\def\ft#1#2{{\textstyle{\frac{\scriptstyle #1}{\scriptstyle #2} } }}
\def\fft#1#2{{\frac{#1}{#2}}}

\def\0{{\sst{(0)}}}
\def\1{{\sst{(1)}}}
\def\2{{\sst{(2)}}}
\def\3{{\sst{(3)}}}
\def\4{{\sst{(4)}}}
\def\5{{\sst{(5)}}}
\def\6{{\sst{(6)}}}
\def\7{{\sst{(7)}}}
\def\8{{\sst{(8)}}}
\def\sst#1{{\scriptscriptstyle #1}}

\def\del{{\partial}}

\def\cG{{{\cal G}}}

\def\scri{{\mathscr{I}}}

\begin{document}

\begin{flushright}
\hfill { UPR-1302-T\ \ \ MI-TH-1944
}\\
\end{flushright}

\begin{center}
{\large {\bf Gaussian Null Coordinates for Rotating Charged Black Holes 
and Conserved Charges
 }}

\vspace{15pt}
{\large M. Cveti\v c$^{1,2}$, 
              C.N. Pope$^{3,4}$, A. Saha$^{3}$ and A. Satz$^{5}$}

\vspace{15pt}

{\hoch{1}\it Department of Physics and Astronomy,\\
University of Pennsylvania, Philadelphia, PA 19104, USA}

\vspace{10pt}

{\hoch{2}\it Center for Applied Mathematics and Theoretical Physics,\\
University of Maribor, SI2000 Maribor, Slovenia}

\vspace{10pt}

\hoch{3}{\it George P. \& Cynthia Woods Mitchell  Institute
for Fundamental Physics and Astronomy,\\
Texas A\&M University, College Station, TX 77843, USA}

\hoch{4}{\it DAMTP, Centre for Mathematical Sciences,
 Cambridge University,\\  Wilberforce Road, Cambridge CB3 OWA, UK}

\hoch{5}{\it Sarah Lawrence College, Bronxville, NY 10708, USA}

\vspace{10pt}

%\vspace{30pt}

%\underline{ABSTRACT}

\end{center}

%\today

%\vfill {\footnotesize Emails: }

%\thispagestyle{empty}

\begin{abstract}

  Motivated by the study of conserved Aretakis charges for a scalar field
on the horizon of an extremal black hole, we construct the metrics
for certain classes of four-dimensional and five-dimensional extremal
rotating black holes in Gaussian null coordinates.  We obtain these as
expansions in powers of the radial coordinate, up to sufficient order to
be able to compute the Aretakis charges.  The metrics we consider are for
4-charge black holes in four-dimensional STU supergravity (including the
Kerr-Newman black hole in the equal-charge case) and 
the general 3-charge black holes in five-dimensional STU supergravity.
We also investigate the circumstances under which the Aretakis charges
of an extremal black hole can be mapped by conformal inversion of the
metric into Newman-Penrose charges at null infinity.  We show that while
this works for four-dimensional static black holes, a simple radial
inversion fails in 
rotating cases because a necessary conformal symmetry of the massless
scalar equation breaks down.  We also discuss that a massless scalar field
in dimensions higher than four does not have any conserved Newman-Penrose
charge, even in a static asymptotically flat spacetime.

\end{abstract}

\pagebreak

\tableofcontents
\addtocontents{toc}{\protect\setcounter{tocdepth}{2}}

%%%%%%%%%%%%%%%%%%%%%%%%%%%%%%%%%%%%%%%%

%\newpage
%%%%%%%%%%%%%%%%%%%%%%%%%%%%%%%%%%%%%%%%
%\documentclass[preprint,showpacs,preprintnumbers,
%  amsmath,amssymb,nofootinbib]{revtex4}

\def\half{\frac{1}{2}}
\def\ben{\begin{equation}}
\def\bea{\begin{eqnarray}}
\def\een{\end{equation}}
\def\eea{\end{eqnarray}}
\def \bp {{\bf p}}
\def \bv {{\bf v}}
\def \bs {{\bf s}}
\def\bt{{\bf t}}

\def \p {\partial}
\def \cL {{\cal  L}}
\def \cG {{\cal  G}}
\def \cLEG {{\cal LEG}}

\def\ft#1#2{{\textstyle{\frac{\scriptstyle #1}{\scriptstyle #2} } }}
\def\fft#1#2{{\frac{#1}{#2}}}

%%%%%%%%%%%%%%%%%%%%%%%%%%%%%%%%%%%%%%%%%%%%%%%%%%%%%%%%%%%%%%%%%%%%%%%%%%%%%%
\section{Introduction}

  In the last few years there have been many studies that have revealed that
the horizon of an extremal black hole is unstable to small perturbations.  
These may be perturbations of the black hole metric itself, or perturbations
of matter fields propagating in the black hole background.  The simplest
such examples arise by considering the perturbations of a 
scalar field \cite{aretakis1,aretakis2,aretakis3,aretakis4}.  Perturbations
of other fields, including linearised gravity, were considered 
in \cite{lucrea}.  The instabilities
stem from the existence of conserved charges on the future horizon of the 
extremal black hole, which imply that physical perturbations
do not decay at large values of the advanced time $v$.
These conserved charges are known as Aretakis charges.  They exist 
quite generally for any black hole with an extremal horizon, but not for
a non-extremal black hole with a bifurcate horizon.

  General arguments for the existence of Aretakis charges in a black hole
with an extremal horizon can be given, making use of the general
near-horizon form of the metric for such black holes \cite{kunlucrea}.
(The near-horizon metric is given in eqn (\ref{gncform}) below.)  The
metric in this form is written using {\it Gaussian null coordinates} (GNC).
The Aretakis charges can be calculated explicitly for a given 
extremal black hole solution by casting the metric into
the Gaussian null form.  This is straightforward for a simple static example
such as the extremal Reissner-Nordstr\"om solution, but it is rather less 
simple for a stationary metric such as the extremal Kerr solution.  In such
a case one cannot construct an exact expression for the metric in 
Gaussian null form, but fortunately it is sufficient to determine just the
first few orders in a GNC expansion of the metric in powers of the 
radial coordinate measuring distance away from the horizon.  This
procedure was carried out for the Kerr metric in \cite{lucli,li2}.  Essentially,
the method used was to solve the equations for null geodesics, in an 
expansion in powers of the radial distance from the horizon.

  One of the main purposes of the present paper is to cast the metrics
for certain classes of rotating extremal supergravity black holes into the
Gaussian null form, thus allowing one to compute the conserved Aretakis
charges in these spacetimes.  Specifically, we carry out this procedure
for the rotating extremal black holes in four-dimensional STU supergravity
that carry four independent electric charges, and also for the
general 3-charge rotating extremal black holes in five-dimensional
STU supergravity.  Specialisations of these results encompass the
previously-derived expressions for the extremal Kerr metric
in four dimensions \cite{lucli} and the extremal Myers-Perry metric in five 
dimensions \cite{li2}.  The intermediate stages in the calculations necessary
for casting the supergravity black hole solutions into Gaussian null form
are quite involved, but the final results that we obtain, at the order
that is sufficient for calculating the Aretakis charges, are remarkably
simple.

  An intriguing observation, in the case of the
extremal Reissner-Nordstr\"om (ERN) metric \cite{bizfri,lumureta}, 
is that by performing an inversion of the radial coordinate so that
the horizon is mapped into future null infinity $\scri^+$, and then 
extracting an overall conformal factor, the Aretakis charges on the
null horizon of the ERN metric can be mapped into conserved Newman-Penrose
charges at null infinity in the conformally-inverted metric.  In fact,
this conformal inversion of the ERN metric actually maps it into the ERN
metric again, a result that had been obtained many years previously by
Couch and Torrence \cite{coutor}.  

    The mapping of Aretakis charges on
an extremal horizon into Newman-Penrose charges on $\scri^+$ was 
investigated in a more general setting in \cite{godgodpop}. It was
shown that the conformal inversion of a general extremal black hole,
written in Gaussian null coordinates, gives rise to a metric that was
called {\it Weakly asymptotically flat} (WAF) in \cite{godgodpop}.  This
metric approaches Minkowski spacetime at infinity, but with rather 
weaker fall-off conditions than those of an asymptotically flat 
spacetime written in Bondi-Sachs coordinates.  It was shown in
\cite{godgodpop} that Newman-Penrose charges can be computed in the
WAF spacetime obtained by conformal inversion of the original extremal
black hole, and in various static examples the
mapping between Aretakis and Newman-Penrose charges was exhibited. These
generalised the mapping for extremal Reissner-Nordstr\"om that was 
found in \cite{bizfri,lumureta}.  In particular, in the more general examples,
such as multi-charge static extremal black holes in STU supergravity,
the conformal inversion of the original metric does not give back the
same metric again, unlike the ERN case.  

   The discussion in \cite{godgodpop} in principle applied also to
stationary extremal black holes that are not static.  One can certainly 
again compute Newman-Penrose charges in the WAF metric obtained by
conformal inversion.  As we shall discuss in the present paper, however, 
in the stationary case 
there is a lacuna in the argument that would be needed in order to
link the Newman-Penrose charges to the original Aretakis charges.  Namely,
in order to map the one into the other, it is necessary to be able to
argue that the solutions of the massless scalar wave equation in the
original black hole metric and in the conformally inverted WAF metric
can be related by the necessary conformal transformation.  This is fine
as long as the Ricci scalar, which enters in the conformally-invariant
scalar equation $(\square -\ft16 R)\psi=0$, either is zero or else it
goes to zero sufficiently rapidly in the asymptotic region.  This is
certainly true in the case of the extremal Reissner-Nordstr\"om solution,
where $R$ vanishes, and also in the more complicated four-dimensional
static supergravity black holes, where $R$ vanishes sufficiently
rapidly asymptotically.  But, as we show later, in the weakened fall-off
of the WAF metrics in the stationary case, the Ricci scalar does not
fall off fast enough at infinity, and this provides an obstruction
to being able to relate the Aretakis charges to the Newman-Penrose charges 
of the conformally-inverted WAF metric, at least if we assume a simple 
inversion of the radial coordinate.  In fact a manifestation of this
problem was foreshadowed in the results in \cite{cvetsatz}, where the
Aretakis and Newman-Penrose charges were calculated in the case of the
extremal Kerr black hole and its conformal inversion.
  
  The focus in \cite{godgodpop}, concerning the relation between
Aretakis charges and Newman-Penrose charges, was four-dimensional 
spacetimes.  We also address in this paper the possible 
extension of these considerations to more than four dimensions.  We
show that conserved Aretakis charges for massless scalar fields exist
in higher-dimensional extremal black holes also.  However, as we
show, Newman-Penrose charges for a massless scalar field 
no longer exist when one goes beyond four 
dimensions.  This happens simply because there is a term in the large-$r$
expansion for the scalar equation written in Bondi-Sachs coordinates
that presents an obstruction to the existence of conserved charges, and
this term occurs with a dimension-dependent coefficient $(n-2)(n-4)$
that is absent in $n=4$ dimensions but not when $n\ge 5$.

\section{General formalism}

  The metric near the horizon of an extremal black hole in any dimension
$n$ can always be written
in Gaussian null coordinates, where it takes the form \cite{kunlucrea}
%%%%%
\be
ds^2= L(x)^2\, \Big[ -\rho^2\, F\, dv^2 + 2 dv d\rho\Big] +
 \gamma_{IJ}\, (dx^I-\rho\, h^I\, dv)(dx^J-\rho\, h^J\, dv)\,,\label{gncform}
\ee
%%%%%
where $F$, $\gamma_{IJ}$ and $h^I$ depend on the radial coordinate $\rho$ and
on the coordinates $x^I$ on the (spherical) horizon, which is located at
$\rho=0$.  Near the horizon we may assume\footnote{In principle it would
suffice to assume weaker asymptotic conditions on the metric functions as 
$\rho$ approaches zero (as discussed, for example, in \cite{godgodpop}), 
but in practice these are the ones that arise in the
black holes we shall be considering.}
%%%%%
\bea
F(\rho,x)&=& 1 + \rho\, F_1(x) + \rho^2\, F_2(x) +\cdots  \,,\nn\\
\gamma_{IJ}(\rho,x) &=& \bar\gamma_{IJ}(x) + \rho\,\gamma^{(1)}_{IJ}(x)
+ \rho^2\, \gamma^{(2)}_{IJ}(x) +\cdots\,,\nn\\
h^I(\rho,x) &=& h^I_0(x) + \rho\, h_1^I(x) + \rho^2\, h_2^I(x)+\cdots\,.
\eea
%%%%% 

   We shall consider the Aretakis charges for a scalar field $\psi$ obeying
the massless Klein-Gordon equation $\square\psi=0$.  The solutions can be
taken to have the small-$\rho$ expansion
%%%%%
\be
\psi(\rho,v,x)= \psi_0(v,x) + \rho\, \psi_1(v,x) + \rho^2\, \psi_2(v,x)+
\cdots\,.
\ee
%%%%%
From this, and the form of the metric expansion, it follows that on
the horizon one has \cite{godgodpop}
%%%%%
\be
\fft{\del}{\del v}\, \Big[2\fft{\del\psi}{\del\rho} +\fft12
 \psi\, \fft{\del\log\gamma}{\del\rho}\Big]+ h_0^I\, \del_I\psi  +
  \fft{1}{\sqrt{\bar\gamma}}\, \del_I\Big(\sqrt{\bar\gamma}\, L^2\, 
\bar\gamma^{IJ}\, \del_J\psi\Big)=0\,.
\ee
%%%%%
If this is integrated over the horizon, with measure 
$\sqrt{\bar\gamma}\, d^{n-2}x$, the final term gives zero since it is a
total derivative. The coordinates $x^I$ on the $(n-2)$-sphere horizon
divide into azimuthal angular coordinates (like $\varphi$ on the 2-sphere),
and latitude type coordinates (like $\theta$ on the 2-sphere).  The
azimuthal coordinates are associated with Killing vectors.  Crucially,
for the extremal 
black hole metrics we shall be considering and as we shall see in detail
later, Gaussian null coordinates 
can be chosen so that $h_0^I$ is zero for
the index values $I$ corresponding to the latitude type coordinates.  It
follows that $h_0^I\,\del_I\psi$ can be written as $(\sqrt{\bar\gamma})^{-1}
\del_I(\sqrt{\bar\gamma}\, h_0^I\, \psi)$, and thus this term is also
a total derivative that integrates to zero.  The upshot is that the quantity
%%%%%
\be
Q_A =\int\sqrt{\bar\gamma}\, d^{n-2} x\, \Big[2\fft{\del\psi}{\del\rho} +\fft12
 \psi\, \fft{\del\log\gamma}{\del\rho}\Big]\,,\label{Qaretakis}
\ee
%%%%%
known as the Aretakis charge,
is conserved on the horizon, in the sense that $\del_v Q_A=0$. 

In the subsequent sections we shall calculate the Aretakis charge for
the case of certain rotating charged extremal black holes in four-dimensional
STU supergravity, and for the general 3-charge extremal rotating black
holes in five-dimensional STU supergravity.  The key part of the calculations
involves constructing the expressions for the black hole solutions in Gaussian
null coordinates, up to the necessary order in the expansion in 
powers of $\rho$.

 A technique for constructing Gaussian null coordinates for
an extremal black hole metric has been described in \cite{lucli,li2}.
Essentially, one writes down the equations for null geodesics in the
extremal metric, for which first integrals exist for the time and the
azimuthal coordinate(s).  The equations for the remaining coordinates
cannot be integrated explicitly, so one then expands these in power
series in the affine parameter $\lambda$ along the geodesics.  The
geodesic equations are then integrated order by order in $\lambda$, imposing
certain transversality conditions in the process.  Finally,
a change of variable from $\lambda$ to $\rho$ brings the metric into the
desired form (\ref{gncform}).

\section{Extremal rotating STU black holes in four dimensions}
\label{sec:4ch4dim}

\subsection{4-charge STU black holes}

The 4-charge four-dimensional STU supergravity black holes that we shall be
considering here were constructed in \cite{cvetyoum}.
A convenient presentation for our purposes can be found in
\cite{chcvlupo}.  The metric can be written as
%%%%%
\bea
ds^2 &=& -\fft{\bar\rho^2 -2 m r}{W}\, (dt + {\cal B}_\1 )^2 +W\, 
\Big(\fft{dr^2}{\Delta} + d\theta^2 + \fft{\Delta\, 
  \sin^2\theta\, d\tilde\phi^2}{\bar\rho^2-2mr}\Big)\,,\label{4chargemet}\\
{\cal B}_\1 &=& \fft{2m a \sin^2\theta\, (r\, \Pi_c - (r-2m)\, \Pi_s)}{
\bar\rho^2 -2m r}\, d\tilde\phi\,,\qquad \bar\rho^2=r^2 + a^2\, \cos^2\theta\,,
\qquad \Delta= r^2 -2m r + a^2\,,\nn\\
W^2 &=& R_1 R_2 R_3 R_4 + a^4\, \cos^4\theta \nn\\
&&+
\Big[ 2r^2 + 2mr \sum_i s_i^2 + 8m^2\, (\Pi_c-\Pi_s)\, \Pi_s\nn\\
&&\qquad -
   4m^2\, (s_1^2 s_2^2 s_3^2 + s_1^2 s_2^2 s_4^2 + s_1^2 s_3^2 s_4^2
             +s_2^2 s_3^2 s_4^2)\Big]\,a^2\, 
\cos^2\theta\,,\nn\\
R_i&=& r+ 2 m s_i^2\,,\qquad \Pi_c=\prod_i c_i\,,\qquad \Pi_s=\prod_i s_i\,,\nn
\eea
%%%%%
where $s_i=\sinh\delta_i$ and $c_i=\cosh\delta_i$.  
The physical mass $M$ and the four physical charges $Q_i$ 
are given by \footnote{In the case where the charges are set equal, with
$\delta_i=\delta$, the solution reduces to the Kerr-Newman black hole,
with $\bar r=r+2m\sinh^2\delta$ being the standard Kerr-Newman 
radial coordinate.  We shall
discuss the extremal Kerr-Newman metric in Gaussian null coordinates in the
next subsection.}
%%%%%
\be
M= m + \ft12 m\, \sum_i s_i^2\,,\qquad
    Q_i=2m s_i\, c_i= m \sinh2\delta_i\,.\label{MQi}
\ee
%%%%%
The metric is extremal when $m=a$, which we
assume from now on.  This implies $\Delta=(r-a)^2$.  
Defining
a new azimuthal coordinate
%%%%%
\be
\phi= \tilde \phi -\fft1{2a\, (\Pi_c+\Pi_s)}\, t
\ee
%%%%%
so that $\del/\del t$ is the Killing vector that becomes null on the 
horizon, we may write the extremal metric in the form
%%%%%
\be
ds^2= W^{-1}\, (A dt^2 + 2B dt d\phi + C d\phi^2) +
   W\Big[ \fft{dr^2}{(r-a)^2} + \fft{du^2}{1-u^2}\Big]\,,\label{4chmet}
\ee
%%%%%
where $u=\cos\theta$.

  In algebraic computations the parameterisation of the charges in terms of 
the four parameters $s_i$ is not ideal, since the relation to the $c_i$
involves square roots, namely $c_i=\sqrt{1+s_i^2}$.  We have found it more
convenient here to work instead with the five parameters $(\alpha,\beta,\gamma,
\Pi_c,\Pi_s)$, where
%%%%%
\bea
\alpha&=& s_1^2 + s_2^2 + s_3^2 + s_4^2\,,\qquad
\beta= s_1^2 s_2^2 + s_1^2 s_3^2 +s_1^2 s_4^2 +
  s_2^2 s_3^2 +s_2^2 s_4^2 + s_3^2 s_4^2\,,\nn\\
\gamma&=& s_1^2 s_2^2 s_3^2 + s_1^2 s_2^2 s_4^2 + s_1^2 s_3^2 s_4^2 +
  s_2^2 s_3^2 s_4^2\,,\qquad
\Pi_c=c_1 c_2 c_3 c_4\,,\qquad \Pi_s= s_1 s_2 s_3 s_4\,.
\eea
%%%%%
These are related by the identity
%%%%%
\be
\gamma= \Pi_c^2 -\Pi_s^2 -1 -\alpha -\beta\,,
\ee
%%%%%
and in practice we find it most convenient to use 
$(\alpha,\beta,\Pi_c,\Pi_s)$ as the four independent 
quantities that parameterise the four
charges.  In terms of $(\alpha,\beta,\gamma,\Pi_c,\Pi_s)$ 
the charge-dependent quantities $W$ and ${\cal B}_\1$ in the metric
(\ref{4chargemet}), subject to the extremality condition $m=a$, are given by
%%%%%
\bea
W^2 &=& r^4 + 2 a \alpha\, r^3 + 2a^2(2\beta + x^2)\, r^2 +
  2 a^3(4\gamma+x^2)\, r +
a^4\, [16\Pi_s^2 + 4(2\Pi_c\Pi_s -2\Pi_s^2 -\gamma) u^2  + u^4]\,,\nn\\
{\cal B}_\1 &=& \fft{2 a^2 \sin^2\theta\, (r\, \Pi_c - (r-2a)\, \Pi_s)}{
\bar\rho^2 -2a r}\, d\tilde\phi\,.
\eea
%%%%%

  Solving now for the null geodesics in this geometry, the first integrals 
for the 
ignorable coordinates $t$ and $\phi$ are taken to be
%%%%%
\be
\dot t= -\fft{C}{W\, (r-a)^2\, (1-u^2)}\,,\qquad
\dot\phi= \fft{B}{W\, (r-a)^2\, (1-u^2)}\,,\label{4chtphidot}
\ee
%%%%%
where a dot denotes a derivative with respect to the affine parameter 
$\lambda$.  We now expand the $r$ and $u$ coordinates as power series in
$\lambda$, with
%%%%%
\be
r=a + \sum_{n\ge 1} R_n(y)\, \lambda^n\,,\qquad
 u= y + \sum_{n\ge 2} X_n(y)\, \lambda^n\,,\label{4ruexp}
\ee
%%%%%
where the affine parameter $\lambda$ vanishes on the horizon.  Substituting
these expansions in to null geodesic constraint 
$g_{\mu\nu} \dot x^\mu\dot x^\nu=0$ and the Euler-Lagrange equation for $u$
allows us to solve iteratively for the $R_n(y)$ and $X_n(y)$ coefficients
in (\ref{4ruexp}).  We find
%%%%%
\bea
R_1(y) &=& \fft{2(\Pi_c+\Pi_s)}{s(y)}\,,\qquad
X_2(y) = \fft{y(1-y^2}{2 a^2 \, s(y)^2}\,,\\
R_2(y) &=&\fft{2(\Pi_c+\Pi_s)(1-y^2)\, [4(\Pi_c-\Pi_s)^3 +
  (9+7\alpha+4\beta) \Pi_s -(5+5\alpha+4\beta) \Pi_c
-(\Pi_c-\Pi_s) y^2]}{a \, s(y)^4}\,,\nn
\eea
%%%%%%
where
%%%%%
\bea
s(y)^2&=& 1+2\alpha+4\beta+8\gamma +16\Pi_s^2 
+2(1+\alpha-2\gamma+4\Pi_c\Pi_s-4\Pi_s^2)\, y^2 + y^4\,,\nn\\
&=& \prod_i (1+2 s_i^2) + 2\Big(1+\sum_i s_i^2 -
   2\sum_{i<j<k} s_i^2 s_j^2 s_k^2 +4\Pi_c\Pi_s -4\Pi_s^2\Big) y^2 + y^4\,.
\label{ssol}
\eea
%%%%%
To the order in $\lambda$ we are working, the coefficient $R_3(y)$ is
needed also.  We shall not present this here, since it is rather complicated.

   The expansions (\ref{4ruexp}) can now be used in (\ref{4chtphidot})
yielding, after integration with respect to $\lambda$,
%%%%%
\be
t= v + T(y,\lambda) + f_t(y)\,,\qquad
\phi= \chi + \Phi(y,\lambda) + f_\phi(y)\,.\label{tphisol}
\ee
%%%%%
Here $T(y,\lambda)$ and $\Phi(y,\lambda)$ are the ``naive'' integrals
of (\ref{4chtphidot}) with respect to $\lambda$. 
The functions of integration $f_t(y)$ and $f_\phi(y)$ are determined
by requiring $V\cdot\del_y=0$, and $(\del_y\cdot\del_\chi)_{\lambda=0}=0$.
Substituting (\ref{4ruexp}) and (\ref{tphisol}) into the
metric (\ref{4chmet}), and working up to and including linear
order in $\lambda$, we obtain the extremal 4-charge metrics
in Gaussian null coordinates $(v,\lambda,y,\chi)$.\footnote{Knowing
just the expansion coefficients $R_1$, $R_2$, $R_3$ and $X_2$ in 
(\ref{4ruexp}) is sufficient to calculate all the metric components
up to and including linear order in $\lambda$, except a term
involving $\lambda \,d\lambda^2$.  Showing that this term is actually
zero would require knowing also the coefficients $R_4$ and $X_3$, in the
procedure we have just described.  However, a simple argument shows that
this term is actually absent, since $g_{\lambda\lambda}= \dot x^\mu \dot x^\nu
\, g_{\mu\nu}$, which vanishes by the null geodesic constraint.}

  Finally, we make a further coordinate 
transformation from $\lambda$ to $\rho$,  of the form
%%%%%
\be
\lambda= w(y)\, \rho\,.\label{lamwrho}
\ee
%%%%%
The function $w(y)$ is determined by the requirement that the metric be
expressible in the form (\ref{gncform}) with, in particular, the function
$F(\rho,x^I)$ being equal to 1 at $\rho=0$.  This gives a first-order
non-linear differential equation for $w$. If we define a new function
$u(y)$ such that $w(y) =a^2\, s(y)\, [\ft12 - u(y)]$ then we find that
$u$ must satisfy
%%%%%
\be
(1-y^2)\, {u'}^2 + u^2 -\ft14 =0\,,
\ee
%%%%%
and so in terms of a new angular coordinate $\vartheta$ such that
$y=\cos\vartheta$, the solutions for $u$ are
%%%%%
\be
u=\ft12 \,,\qquad \hbox{or}\qquad u=-\ft12\,,\qquad \hbox{or}\qquad
u=\ft12 \sin(\vartheta+k)\,,\label{usols}
\ee
%%%%%
where $k$ is an arbitrary constant of integration. The metric functions
$L(x)$ and $h_0^I$ in the metric (\ref{gncform}) are then given by
%%%%%
\be
L^2= w(y)\,,\qquad h_0^y= (1-y^2)\, u'(y)\,,
\qquad h_0^\chi= -\fft{(\Pi_c-\Pi_s)}{(\Pi_c+\Pi_s)}\, [\ft12-u(y)]\,.
\ee
%%%%%
In order for the metric (\ref{gncform}) to be non-degenerate $L(x)^2$, and hence
$w(y)$, should be positive everywhere on the sphere. The first solution in
(\ref{usols}) is
trivial, implying $w(y)=0$ and hence $L=0$, so this is excluded.  The 
second solution gives
%%%%%
\be
w(y)=a^2\, s(y)\,,\qquad h_0^y=0\,,\qquad
h_0^\chi = -\fft{(\Pi_c-\Pi_s)}{(\Pi_c+\Pi_s)}\,,
\ee
%%%%%
and so $w(y)$ is indeed positive everywhere on the sphere.  Note that
$h_0^y=0$ for this solution.  As we noted previously when discussing the
derivation of the Aretakis charge, it is necessary that $h_0^y$ vanish in
order for the charge to be conserved on the horizon.
The third solution in (\ref{usols}) gives
%%%%%
\bea
w(y) &=&\ft12 a^2\,s(y)\, [1-\sin(\vartheta+k)]\,,\nn\\
\qquad h_0^y&=&-\sin\vartheta\, \cos(\vartheta+k)\,,
\qquad h_0^\chi= -\fft{(\Pi_c-\Pi_s)}{2(\Pi_c+\Pi_s)} \, 
[1-\sin(\vartheta+k)] \,.
\eea
%%%%%
The function $w(y)$ will be positive everywhere on the sphere provided
that $k$ is chosen appropriately.  However, $h_0^y$ is non-zero, and so there
will be no conserved Aretakis charge in this case.  Thus we are led to choose
the second solution in (\ref{usols}), implying that
%%%%%
\be
\lambda=a^2\, s(y)\, \rho\,,
\ee
%%%%%
where $s(y)$ is given in (\ref{ssol}).
The metric then takes the form (\ref{gncform}) with
%%%%%
\be
L(y)^2 = a^2\, s(y)\,,\qquad F= 1 -2 a (\Pi_c-\Pi_s)\, \rho +\cdots\,,
\label{LFsol}
\ee
%%%%%
and
%%%%%
\bea
h_0^y &=& 0\,,\qquad h_0^\chi = -\fft{(\Pi_c-\Pi_s)}{\Pi_c + \Pi_s}\,,\nn\\
\bar\gamma_{yy} &=& \fft{a^2\, s(y)}{1-y^2}\,,\qquad 
\bar\gamma_{\chi\chi}= \fft{4 a^2\, (\Pi_c+\Pi_s)^2\, (1-y^2)}{s(y)}\,,
\qquad \bar\gamma_{y\chi}=0\,,\nn\\
\gamma_{yy}^{(1)}&=& \fft{2a^3 (\Pi_c+\Pi_s)[ 4\Pi_c^2-4\Pi_s^2 -
  (2+\alpha)(1-y^2) ]}{(1-y^2)s(y)}\, \,,\nn\\
\gamma_{y\chi}^{(1)} &=&
\fft{4 a^3 (\Pi_c+\Pi_s)\, y\, (1-y^2)}{s(y)}\,,\label{4chgncmet}
\\
\gamma_{\chi\chi}^{(1)} &=& \fft{8a^3\, (\Pi_c+\Pi_s)^2\, (1-y^2)}{s(y)^4}\,
\Big[ (\Pi_c+\Pi_s)\Big( (2+\alpha)(1-y^2) -4\Pi_c^2 + 4\Pi_s^2\Big)
  +   2(\Pi_c-\Pi_s) s(y)\Big]\,,\nn
\eea
%%%%%
We have calculated $h_1^y$ and $h_1^\chi$ but they are rather complicated,
and we shall not present them here since they are not required for our 
subsequent purposes.  

  It is now a straightforward matter to calculate the Aretakis charge
(\ref{Qaretakis}) for the scalar field in this extremal black hole 
From
(\ref{4chgncmet}) the
evaluation of $\del\log\gamma/\del \rho$ on the horizon gives
%%%%%
\be
\fft{\del \log\gamma}{\del\rho}\Big|_{\rho=0} =
 4a\, (\Pi_c-\Pi_s)\,,
\ee
%%%%%
and hence the Aretakis charge (\ref{Qaretakis}) is given by
%%%%%
\be
Q_A =2a^2\, (\Pi_c+\Pi_s)\int dy d\chi\, \Big[2\fft{\del\psi}{\del\rho} +
 2 a\, (\Pi_c-\Pi_s)\psi\Big],\label{Qaretakis4}
\ee
%%%%%
evaluated at $\rho=0$, i.e.
%%%%%
\be
Q_A =2a^2\, (\Pi_c+\Pi_s)\int dy d\chi\, \Big[2\psi_1  +
 2 a\, (\Pi_c-\Pi_s)\psi_0\Big],\label{Qaretakis4b}
\ee
%%%%%

\subsection{Kerr-Newman black holes}

  In the case where the four electric charges in the STU black hole
are set equal, by taking $\delta_i=\delta$, and hence $s_i=s$, $c_i=c$,
one obtains the Kerr-Newman black hole.  From (\ref{MQi}) the physical
mass $M$ and electric charge $Q=Q_i$ become
%%%%%
\be
M=m(1+2s^2)\,,\qquad Q= 2msc\,,
\ee
%%%%%
and so the extremality condition $a=m$ can be written as
%%%%%
\be
M^2=a^2+Q^2\,.
\ee
%%%%%
The function $s(y)$ in (\ref{ssol}) is now polynomial in $y$, given by
%%%%%
\be 
a^2\, s(y)= M^2 + a^2\, y^2\,,
\ee
and the metric coefficients in (\ref{LFsol}) and (\ref{4chgncmet}) become
%%%%%
\be
L(y)^2 = M^2 + a^2\, y^2\,,\qquad F=1-2M \rho+\cdots\,,
\ee
%%%%%
and 
%%%%%
\bea
h_0^y&=&0\,,\qquad h_1^y= -\fft{2 a^2\, M\, (a^2+M^2)\, y\, (1-y^2)}{
    (M^2+a^2 \, y^2)^2}\,,\nn\\
h_0^\chi &=& -\fft{2 a M}{a^2+M^2}\,,\qquad
h_1^\chi= \fft{a\, [2 M^4 + a^2\, M^2\, (1-y^2)\, -a^4\, y^2\, (1+y^2)]}{
  (a^2+M^2)(M^2+a^2\, y^2)}\,,\nn\\
\gamma^{(0)}_{yy} &=& \fft{M^2+a^2\, y^2}{1-y^2}\,,\qquad
\gamma^{(0)}_{\chi\chi}=\fft{(a^2+M^2)^2\,(1-y^2)}{M^2 + a^2\, y^2}\,,
\qquad \gamma^{(0)}_{y\chi}=0\,,\nn\\
\gamma^{(1)}_{yy}&=& \fft{2 M\, (a^2+M^2)}{(1-y^2)}
\,,\quad
\gamma_{\chi\chi}^{(1)} =
\fft{2 M (a^2+M^2)^2\, (1-y^2)[M^2-a^2(1-2y^2)]}{
   (M^2+a^2\, y^2)^2}\,,\label{gamma1n}\\
\gamma_{y\chi}^{(1)} &=&
\fft{2 a^3\, (a^2+M^2)\, y\, (1-y^2)}{(M^2+a^2\, y^2)}\,.
\eea
%%%%%
The Aretakis charge (\ref{Qaretakis4b}) becomes
%%%%%
\be 
Q_A = (a^2+M^2)\int dy d\chi\, (2\psi_1 + M \psi_0)\,.
\ee
%%%%%

  It is worth remarking that in the case of pairwise-equal charges, where
we take, for example, $\delta_3=\delta_1$ and $\delta_4=\delta_2$, the 
function $s(y)$ given in (\ref{ssol}) again becomes purely polynomial in $y$,
with
%%%%%
\be 
s(y)= (1+2s_1^2)(1+2s_2^2) + y^2\,,
\ee
%%%%%
and, as can straightforwardly be seen, 
the expressions for the metric coefficients in
(\ref{LFsol}) and (\ref{4chgncmet}) again simplify considerably.  The
extremality condition $a=m$, expressed in terms of the physical
mass and charges, is now
%%%%%
\be
a^2= \fft{(2M-Q_1-Q_2)(2M-Q_1+Q_2)(2M+Q_1-Q_2)(2M+Q_1+Q_2)}{16 M^2}\,.
\ee
%%%%%

\section{Extremal rotating STU black holes in five dimensions} 

  The metric for the general 3-charge rotating black holes 
of five-dimensional STU supergravity was constructed in \cite{cvetyou5}.  
It is convenient to introduce the 
quantities $\alpha$, $\beta$, $\Pi_s$ and $\Pi_c$, defined in terms of
the boost parameters $\delta_i$ for $i=1$, 2 and 3 by
%%%%%
\be
\alpha=s_1^2+s_2^2+s_3^2\,,\qquad \beta=s_1^2 s_2^2 + s_1^2 s_3^2
+ s_2^2 s_3^2\,,\qquad \Pi_s= s_1 s_2 s_3\,,\qquad \Pi_c= c_1 c_2 c_3\,,
\label{3repboost}
\ee
%%%%%
where $s_i=\sinh\delta_i$ and $c_i=\cosh\delta_i$.  They obey the relation
%%%%%
\be
\Pi_c^2 = 1+\alpha+\beta +\Pi_s^2\,,
\ee
%%%%%
which we employ in order to eliminate $\beta$.  We use the coordinates
$t$, $r$, $\theta$, $\tilde \phi$ and $\widetilde\psi$ of \cite{cvetyou5},
with the redefinition $u=\cos\theta$.  (We have put tildes one the
two azimuthal angles because will shortly redefine untilded versions 
with respect to which the null vector on the horizon is simply
$\del/\del t$.)  Extremality of the metric is achieved 
by taking mass parameter $\mu=(a+b)^2$, where $a$ and $b$ are the
two rotation parameters (called $l_1$ and $l_2$ in \cite{cvetyou5}); 
the double horizon is then at $r=\sqrt{ab}$.
The metric can be written as
%%%%%
\bea
ds^2 &=& \Delta^{-2/3} \, (A\, dt^2 + 2 B_1\, dt d\phi + 2 B_2\, dt d\psi +
         C_1 \,d\phi^2 + C_2 \,d\psi^2 + 2 C_{12} \,d\phi d\psi)\nn\\
&& +
    \Delta^{1/3} \Big[ \fft{dr^2}{(r^2-ab)^2} +\fft{du^2}{1-u^2}
\Big]\,,
\eea
%%%%%
where
%%%%%
\be
\Delta= (a+b)^6\, \Pi_s^2 + (a+b)^4\, \beta\, \rho^2 + (a+b)^2\, \alpha\, \rho^4
    + \rho^6\,,
\ee
%%%% 
and $\rho^2 = r^2+ a^2\, u^2 + b^2\, (1-u^2)$.
The coordinate redefinitions for the untilded azimuthal coordinates are
%%%%%
\be
\phi=\tilde\phi - \fft{1}{(a+b)(\Pi_c+\Pi_s)}\, t\,,\qquad
\psi=\widetilde\psi - \fft{1}{(a+b)(\Pi_c+\Pi_s)}\, t\,.
\ee
%%%%%
The functions $A$, $B_1$, $B_2$, $C_1$, $C_2$ and $C_{12}$ 
can be read off by starting
from the metric given in eqn (18) of \cite{cvetyou5} and applying the steps
detailed above.

  We proceed along the same lines as in the previous four-dimensional examples,
with the appropriate generalisations to five dimensions.  The case of
the uncharged five-dimensional Myers-Perry black hole is discussed in 
detail in \cite{li2}.  The first integrals for the
geodesics in the $t$, $\phi$ and $\psi$ directions are chosen so that
%%%%%
\be
A \,\dot t + B_1 \,\dot\phi + B_2 \,\dot\psi= \Delta^{2/3}\,,\quad
B_1\, \dot t+ C_1\, \dot\phi + C_{12}\,\dot\psi=0\,,\quad
B_2\, \dot t+ C_2 \,\dot\psi + C_{12}\,\dot \phi=0\,.\label{tphipsidot}
\ee
%%%%%
We then make the expansions
%%%%%
\be
r^2 = ab + \sum_{n\ge 1} R_n(y)\, \lambda^n\,,\qquad
u= y + \sum_{n\ge 2} X_n(y)\, \lambda^n\,,
\ee
%%%%%
where $\lambda$ is the affine parameter along the null geodesic.
Plugging these into the null constraint equation $L\equiv\ft12 g_{\mu\nu}\,
\dot x^\mu \dot x^\nu=0$ and the second-order Euler-Lagrange 
equation for $u$ allows us to solve iteratively for the $R_n(y)$ and
$X_n(y)$ coefficient functions.  We find
%%%%%
\bea
R_1(y)&=& \fft{2\sqrt{ab}\, (a+b)(\Pi_c+\Pi_s)}{s(y)}\,,
\qquad
X_2(y)= \fft{(a-b) y (1-y^2)}{2(a+b) s(y)^2}\,,\nn\\
R_2(y)&=& \fft{(a+b)(\Pi_c+\Pi_s)}{3s(y)^5}\, \Big[ 3(a+b)(\Pi_c-\Pi_s) s(y)^3
  - 2 ab (\Pi_c+\Pi_s)F_1\Big]\,,
\eea
%%%%%
where $s(y)$ is given by
%%%%%%
\be
s(y)= \prod_{i=1}^3 [ay^2+b(1-y^2) +(a+b)s_i^2\,]^{1/3}\,,
\ee
%%%%%
and $F_1$ is given in (\ref{Fexp}) below.
To the order in $\lambda$ that we are working, we also need $R_3(y)$.
We have calculated this but it is too complicated to present explicitly here.
Substituting these results into eqns (\ref{tphipsidot}) enables us to
solve for $\dot t$, $\dot\phi$ and $\dot\psi$.  After integration, we
have
%%%%%
\be
t= v + T(y,\lambda) + f_t(y)\,,\qquad
\phi= \chi + \Phi(y,\lambda) + f_\phi(y)\,,\qquad
\psi= \sigma + \Psi(y,\lambda) + f_\psi(y)\,,
\ee
%%%%%
where $T$, $\Phi$ and $\Psi$ are the ``naive'' $\lambda$ integrals, as
discussed in the four-dimensional examples previously.  The functions of
integration $f_t(y)$, $f_\phi(y)$ and $f_\psi(y)$ are determined by 
requiring $V\cdot \del_y=0$, where $V=\dot x^\mu \del_\mu$,
along with $(\del_y\cdot\del_\chi)_{\lambda=0}=0$
and $(\del_y\cdot\del_\sigma)_{\lambda=0}=0$.  

  With these preliminaries, we now have the necessary coordinate 
redefinitions to re-express the metric in terms of the Gaussian 
null coordinates $(v,\lambda,y,\chi,\sigma)$.  A coordinate
transformation 
%%%%%
\be
\lambda= (a+b)\, s(y)\, [\ft18 - u(y)]\, \rho\label{lamrho5}
\ee
%%%%%
then casts it into the form (\ref{gncform}), with $F(\rho,x^I)=1$ at $\rho=0$,
provided that $u(y)$ obeys
%%%%%
\be
y(1-y)\, {u'}^2 + u^2 - \ft1{64}=0\,.
\ee
%%%%%
Letting $y=\ft12(1+\cos\vartheta)$, this implies
%%%%%
\be
u=\ft18\,,\qquad\hbox{or}\qquad u=-\ft18\,,\qquad\hbox{or}\qquad
u=\ft18 \sin(\vartheta+k)\,,
\ee
%%%%%
where $k$ is an arbitrary constant.  The metric function $h^0_y$ is given
by $h^0_y=4y(1-y)\, du/dy$, and thus we have just the one solution $u=-\ft18$
that gives both a non-singular coordinate transformation (\ref{lamrho5})
and a vanishing $h^0_y$ (as is required for obtaining an Aretakis charge).
The final form of the five-dimensional 3-charge extremal STU black holes in
Gaussian null coordinates is then given by (\ref{gncform}) with
%%%%%
\bea
L(y)^2 &=& \ft14(a+b)\, s(y)\,\qquad
F= 1-\fft{(a+b)^2\,(\Pi_c-\Pi_s)}{4\sqrt{ab}}\, \rho+\cdots\,,\nn\\
h_0^y &=&0\,,\qquad h_0^\chi = -\fft{(b\, \Pi_c-a\, \Pi_s)}{2\sqrt{ab}\, 
(\Pi_c+\Pi_s)}\,,\qquad
h_0^\sigma = -\fft{(a\, \Pi_c-b\, \Pi_s)}{2\sqrt{ab}\,
(\Pi_c+\Pi_s)}\,,\nn\\
\bar\gamma_{yy} &=& \fft{(a+b)\, s(y)}{1-y^2}\,,\qquad
  \bar\gamma_{\chi\chi} = \fft{(a+b)\, H_2\, (1-y^2)}{s(y)^2}\,,\qquad
\bar\gamma_{\sigma\sigma}=\fft{(a+b)\, \widetilde H_2\, y^2}{s(y)^2}\,,
\nn\\
\bar\gamma_{\chi\sigma} &=& \fft{(a+b)\,H_0\, y^2\, (1-y^2)}{s(y)^2}\,,
\qquad \gamma_{yy}^{(1)} =
 \fft{\sqrt{ab}(a+b)^2\,(\Pi_c+\Pi_s) H_1}{6(1-y^2) s(y)^2}\,,\nn\\
\gamma_{y\chi}^{(1)}&=& \fft{(a-b)(a+b)^2\, y(1-y^2) H_4}{2s(y)^2}\,,
\qquad \gamma_{y\sigma}^{(1)} =
   \fft{(a-b)(a+b)^2\, y^2 \widetilde H_4}{2s(y)^2}\,,
\nn\\
\gamma_{\chi\chi}^{(1)} &=& -\fft{\sqrt{ab} (a+b)^2\, (\Pi_c+\Pi_s)
  [2 H_1 H_2 - 3 H_3\, s(y)^3](1-y^2)}{6 s(y)^5}\,,\nn\\
\gamma_{\sigma\sigma}^{(1)} &=& -\fft{\sqrt{ab} (a+b)^2\, (\Pi_c+\Pi_s)
  [2 H_1 \widetilde H_2 - 3 \widetilde H_3\, s(y)^3]y^2}{6 s(y)^5}\,,\nn\\
\gamma_{\chi\sigma}^{(1)} &=& -\fft{\sqrt{ab} (a+b)^2\, (\Pi_c+\Pi_s)
  [2 H_0 H_1 - 3 ab\, s(y)^3]y^2(1-y^2)}{6 s(y)^5}\,.\label{5Dgnc}
\eea
%%%%%
Here, the functions $H_a$ and $\widetilde H_a$ are given by
%%%%%
\bea
H_0 &=& (a+b)^3\, \Pi_c \Pi_s + ab^2 [ 1+\alpha -(\Pi_c+\Pi_s)^2 + (1-y^2) ] +
 a^2 b [ 1+\alpha -(\Pi_c+\Pi_s)^2 + y^2 ]\,,\nn\\
H_1 &=& (a+b)^2(1-\Pi_c^2+\Pi_s^2) -3 (a-b)^2 y^2(1-y^2) +
  [a^2(3+\alpha) -\alpha b^2] y^2 +
 [b^2(3+\alpha) -\alpha a^2] (1-y^2)\,,\nn\\
H_2&=& ab(a+b) (\Pi_c+\Pi_s)^2 + (a^3\Pi_c^2+b^3\Pi_s^2) y^2 
\nn\\
&&-ab [a(1+\alpha-\Pi_c^2 + 2\Pi_c \Pi_s) + 
       b(2+\alpha -\Pi_s^2+ 2\Pi_c\Pi_s)]y^2 -ab(a-b) y^4\,,\nn\\
H_3 &=& (a+b)^2 (\Pi_c^2-\Pi_s^2) + [a^2(1+\alpha) -b^2 (2+\alpha)] y^2 +
   (a-b) y^4\,,\nn\\
H_4 &=& ab (\Pi_c+\Pi_s) + (a-b)(a \Pi_c - b \Pi_s) y^2\,,\nn\\
H_5 &=& (a+b)(\Pi_c+\Pi_s)(b\Pi_c-a\Pi_s) + (a-b)[
  a(1+\alpha) + b(2+\alpha)] y^2 + (a-b)^2 y^4\,,\label{Fexp}
\eea
%%%%%
with $(\widetilde H_2, \widetilde H_3,\widetilde H_4,\widetilde H_5)$
being obtained from $(H_2,H_3,H_4,H_5)$ by making the replacements
%%%%%
\be
\widetilde H_a = H_a\Big|_{(y\rightarrow \sqrt{1-y^2}\,, \ a\rightarrow b\,,
\ b\rightarrow a)}\,.
\ee
%%%%%
($H_0$ and $H_1$ are invariant, or ``self dual,'' under this transformation.)
Note that, as in the analogous discussion in the case of the 4-charge
black holes in four dimensions, our choice of coordinate redefinition 
in (\ref{lamrho5}) with $u=-\ft18$ 
has ensured that $h_0^y=0$, which is essential in for
the existence of a conserved Aretakis charge.

 From the expressions in (\ref{5Dgnc}), we find that
%%%%%
\be
\fft{\del \log\gamma}{\del\rho}\Big|_{\rho=0} =\fft{(a+b)^2\, (\Pi_c-\Pi_s)}{
2\sqrt{ab}}
\ee
%%%%%
for the five-dimensional 3-charge rotating extremal black holes, and
$\sqrt{\bar\gamma}= \sqrt{ab}\, (a+b)^2\, (\Pi_c+\Pi_s)\, y$, 
so the Aretakis charge (\ref{Qaretakis}) is given by
%%%%%
\be
Q_A = \sqrt{ab}\, (a+b)^2\, (\Pi_c+\Pi_s)\,\int y dy\, d\chi\,d\sigma\,
\Big[ 2\psi_1 + \fft{(a+b)^2\, (\Pi_c-\Pi_s)}{4\sqrt{ab}}\, \psi_0
\Big]\,.
\ee
%%%%%

\section{Inversion and Newman-Penrose Charges}

\subsection{Inversion and weakly asymptotically flat spacetimes in four 
dimensions}

  Conserved 
Aretakis charges are defined on the horizon of an extremal black hole.
A different kind of conserved charge, known as a Newman-Penrose (NP) charge, is
defined at null infinity in an asymptotically flat spacetime.  It was shown in
\cite{coutor} that in the case of an extremal Reissner-Nordstr\"om (ERN)
black hole, there exists an inversion symmetry, which takes the 
form $\rho\rightarrow
1/r$, where $\rho$ is the radial coordinate of the ERN black hole written
in Gaussian null coordinates, under which the inverted metric is conformally 
related to the ERN metric again.  This symmetry was employed in 
\cite{bizfri,lumureta} in order to show that the Aretakis charge for
a massless scalar field in the ERN background was related to the Newman-Penrose
charge for the massless scalar, calculated at future null infinity in the
same ERN metric.

The fact that the conformal inversion of the extremal Reissner-Nordstr\"om 
black hole gives back precisely the same ERN metric is of itself inessential
for the purpose of mapping the Aretakis charge into a Newman-Penrose
charge in the conformally inverted metric.  In \cite{godgodpop}, a
general discussion was given in which an extremal black hole metric,
written in Gaussian null coordinates as in (\ref{gncform}), was conformally
inverted to give a metric that was {\it weakly asymptotically flat} (WAF).  
Under
appropriate conditions, the Aretakis charge in the original extremal
black hole metric can be mapped into a Newman-Penrose charge in the
weakly asymptotically flat metric.  The conformal inversion is
effected by starting from (\ref{gncform}) and then taking
%%%%%
\be
\rho=\fft1{r}\,,\qquad v=u\,,\qquad ds^2= \fft{L(x)^2}{r^2}\, d\tilde s^2\,,
\ee
%%%%%
and the weakly asymptotically flat metric is thus given by \cite{godgodpop}
%%%%%
\be
d\tilde s^2 = -F\, e^{2\beta}\, du^2 - 2 e^{2\beta}\, du dr +
r^2\, h_{IJ}\, (dx^I-C^I\, du)(dx^J- C^J\, du)\,,\label{WAFform}
\ee
%%%%%
with
%%%%%
\bea
\beta &=&0\,,\qquad 
C^I= \fft{h^I}{r} = h_0^I\, r^{-1} + h_1^I\, r^{-2} +\cdots\,,
\nn\\
 h_{IJ}&=& L(x)^{-2}\, \gamma_{IJ} =L(x)^{-2}\, (\bar\gamma_{IJ} +
\gamma^{(1)}_{IJ}\, r^{-1} +\cdots)\,.
\eea
%%%%%
The appellation ``weakly asymptotically flat'' signifies the fact that
the usual definition of asymptotic flatness has been weakened in two
respects.  Firstly, the vector $C^I$ is allowed to have terms at order 
$1/r$ in its asymptotic expansion, in contrast to the usual requirement
of leading $1/r^2$ fall-off for asymptotic flatness.  Secondly, the
leading-order metric on the $r=\,$constant spatial sections is not restricted
to being that of a round sphere.

In the case of static spherically-symmetric 
extremal black hole metrics in four dimensions, the distinction between 
weakly asymptotically flat and asymptotically flat in the conformally inverted
metric is irrelevant, since $h^I=0$ and hence $C^I=0$, and in addition
$L(x)=\,$ constant and $L(x)^{-2}\, \bar\gamma_{IJ}$ is just 
the metric on the unit round sphere, implying that $h_{IJ}$ at leading order is
also the round sphere metric.  In \cite{godgodpop} the conformal inversion
for static spherically-symmetric extremal black holes was employed in order
to relate the Aretakis and Newman-Penrose charges in some more general
examples, such as the 4-charge extremal static black holes of four-dimensional
STU supergravity.

A crucial point about the conformal inversion in the static four-dimensional
extremal black holes considered in \cite{godgodpop} 
is that the Ricci scalar is either zero (as in the
extremal Reissner-Nordstr\"om example) or else it goes to zero
sufficiently rapidly (as in the general 4-charge STU examples) 
that in both the Aretakis and the Newman-Penrose calculations, one can
as well replace the massless scalar operator $\square$ by the 
conformally invariant operator $\square -\ft16 R$.  This means that one
can invoke the consequent conformal relation between the solutions of
the scalar operator in the original and the conformally inverted metrics, in
order to establish a mapping between the Aretakis and the Newman-Penrose
charges.

If the extremal black hole is stationary but not static, the mapping between
the Aretakis charge in the black hole metric and the Newman-Penrose charge
in the conformally inverted metric with $\rho\rightarrow 1/r$ 
will break down.  Even if we consider
the simplest example, namely the extremal Kerr metric, the conformal
mapping between the solutions of the scalar wave equation will fail.  There
is no problem with the calculation for the
Aretakis charge in the extremal Kerr metric, since the Ricci scalar vanishes
and there is no difference between the massless scalar operator $\square$ and
the conformally invariant operator $\square -\ft16 R$.  However, in
the weakly asymptotically flat metric $d\tilde s^2$ obtained by conformal 
inversion, the Ricci scalar is non-vanishing and it has a $1/r^2$ fall-off
at large $r$.  Thus one can straightforwardly see from the form of the 
four-dimensional extremal rotating 4-charge metrics in Gaussian null
coordinates obtained in section \ref{sec:4ch4dim} that after turning
off the charges to give the Kerr metric, and making the
inversion and conformal scaling
%%%%%
\be
\rho=\fft1{r}\,,\qquad d\tilde s^2 = \fft{r^2}{L(x)^2}\,ds^2\,,
\ee
%%%%%
the Ricci scalar $\widetilde R$ calculated in the conformally-rescaled metric
has the leading-order form
%%%%%
\be
\widetilde R =\fft{6(1-5y^2)}{(1+y^2)^2}\,\fft1{r^2} + 
    {\cal O}\Big(\fft1{r^3}\Big)\,.
\ee
%%%%%
When one looks at the calculation of the Newman-Penrose
charge (which we shall discuss in detail below), one finds that while there
does exist a Newman-Penrose charge for a massless scalar obeying 
$\widetilde\square\widetilde \psi=0$ in the WAF metric, 
there does not exist a Newman-Penrose charge
for a scalar obeying the conformally invariant equation 
$(\widetilde \square-\ft16 \widetilde R)\widetilde \psi=0$.  
The problem is that the Ricci scalar with its 
$1/r^2$ fall-off gives an obstruction to the existence of a conserved
charge.  There is therefore no conformal mapping that allows one to
relate the Aretakis charge for a scalar obeying $\square\psi=0$ in the
extremal Kerr metric to the Newman-Penrose charge for a scalar obeying
$\widetilde\square\widetilde \psi=0$ in the related WAF metric.  
This difficulty can be seen 
in a calculation of the Aretakis and the Newman-Penrose charges for this
example that was carried out in \cite{cvetsatz}.

  It is useful nevertheless to examine in detail the construction of the
Newman-Penrose charge for a massless scalar field in a general weakly
asymptotically flat spacetime, and we shall now present the calculation in
a general spacetime dimension $n$.

\subsection{Inversion and WAF spacetimes in higher dimensions}

  In dimensions greater than four, further complications can arise.
If we consider even a static extremal black hole, such as a
higher-dimensional extremal Reissner-Nordstr\"om metric, the
Ricci scalar no longer vanishes (since only in four dimensions is the
the electromagnetic energy-momentum tensor trace-free), and in fact
it approaches a constant on the horizon.  For example, writing 
the five-dimensional
ERN metric in the Gaussian null form (\ref{gncform}), it is given by
%%%%%
\bea
L(x)^2 &=& \fft{Q}{4}\,,\qquad F= \rho^2\, 
\Big(1+\fft{\sqrt{Q}}{8}\, \rho\Big)^2\,
\Big(1+\fft{\sqrt{Q}}{4}\, \rho\Big)^{-4}\,,\nn\\
\gamma_{IJ}&=& Q\, \Big(1+\fft{\sqrt{Q}}{4}\, \rho\Big)^2\, \omega_{IJ}\,,
\eea
%%%%%
where $Q$ is the electric charge and $\omega_{IJ}$ is the metric on the
unit 3-sphere.  The Ricci scalar is given by
%%%%%
\be
R= -\fft{2}{Q}\, \Big(1+\fft{\sqrt{Q}}{4}\, \rho\Big)^{-6}\,,
\ee
%%%%%
which approaches the constant $-2/Q$ on the horizon at $\rho=0$.  This 
implies that although there exists an Aretakis charge for a scalar
field obeying the massless wave equation $\square\psi=0$, there will
be no conserved charge for a scalar obeying the conformally invariant 
wave equation, which is $(\square -\ft{3}{16}\, R)\psi=0$ in five
dimensions. 

  Furthermore, after the inversion with $\rho\rightarrow 1/r$ and 
$v\rightarrow u$, and the conformal scaling
to $d\tilde s_5^2= (r^2/L(x)^2)\, ds_5^2$, we obtain the WAF metric
%%%%%
\be
d\tilde s_5^2 = -\fft{\Big(1+\fft{\sqrt{Q}}{8r}\Big)^2}{
 \Big(1+\fft{\sqrt{Q}}{4r}\Big)^4}\, du^2 - 2 du dr +
  4r^2\, \Big(1+\fft{\sqrt{Q}}{4r}\Big)^2\,d\Omega_3^2\,,\label{waf5}
\ee
%%%%%
where $d\Omega_3^2$ is the metric on the unit 3-sphere.  This has a 
Ricci scalar that falls off only as $r^{-2}$ at infinity, and in fact
%%%%%
\be
\widetilde R = -\fft{9}{2 r^2} + {\cal O}\Big(\fft1{r^3}\Big)\,.\label{5dR}
\ee
%%%%%
The reason why this happens in higher dimensions but not in four is that
now, in the conformally inverted WAF metric, the metric on the $r=\,$constant
surfaces is not simply approaching $r^2\, d\Omega^2$, but
instead a non-unit constant times $r^2\, d\Omega^2$.  (The constant is 4 in
the five-dimensional example in (\ref{waf5}).)  This means the spatial 
metric is not locally approaching the Euclidean metric at large
$r$, and this is responsible for
the slower falloff of the Ricci scalar.

As discussed previously, a Ricci scalar with this fall-off contributes 
in the calculation of Newman-Penrose charges if one considers the
conformally invariant scalar wave operator.  

  As we shall see in the next section, there is actually a further
complication in dimensions greater than four, when one attempts to
construct conserved Newman-Penrose charges at null infinity.  

\subsection{Newman-Penrose charges in four and higher dimensions}

  In this section, which is concerned exclusively with the 
calculation of Newman-Penrose charges, we shall drop the tildes that
we were previously using to denote the weakly asymptotically flat
metric.
  The general solution of $\square\Psi=0$ in the weakly asymptotically 
flat metric (\ref{WAFform}) in $n$ dimensions has a large-$r$ expansion
of the form
%%%%%
\be
\Psi(r,u,x)=\Psi_0(u,x)\, r^{-\gamma} +\Psi_1(u,x)\, r^{-\gamma-1}+\cdots\,,
\qquad \gamma= \fft{n-2}{2}\,.\label{Psiexp}
\ee
%%%%%
If the $(n-2)$-dimensional metric $h_{IJ}$ in the WAF metric (\ref{WAFform}) 
is expanded as
%%%%%
\be
h_{IJ}(r,u,x) = \omega_{IJ} + h^{(1)}_{IJ}\, r^{-1}+\cdots\,,
\ee
%%%%%
we may choose a coordinate gauge where $\sqrt{h} = \zeta(r)\, \sqrt{\omega}$,
with
%%%%%
\be
\zeta(r) = 1 + \zeta_0\, r^{-1} + \zeta_1\, r^{-2}+\cdots\,.
\ee
%%%%%
Substituting (\ref{Psiexp}) into $\square\Psi=0$, evaluated in the 
WAF metric background, we find that at the leading order in the
large-$r$ expansion, 
%%%%%
\be
\fft{\del}{\del u}\, [2\Psi_1 + \zeta_0\, \Psi_0] - 
\fft{(n-2)(n-4)}{4}\,\Psi_0 -\fft{n-4}{2}\, C_0^I\, \del_I\Psi_0
+\fft{n-2}{2}\, D_I(C_0^I\, \Psi_0)  + D^ID_I\Psi_0=0\,,\label{NPcon}
\ee
%%%%%
where $C^I$ has an expansion of the form 
$C^I= C_0^I\, r^{-1} + C_1^I\, r^{-2}+\cdots$, and where $D_I$ denotes the 
covariant derivative in the $\omega_{IJ}$ metric.

   In $n=4$ dimensions, we can obtain a conserved charge by integrating
(\ref{NPcon}) over the 2-sphere with metric $\omega_{IJ}$:
%%%%%
\be
Q_{NP} = \int \sqrt{\omega}\, d^2x\,  [2\Psi_1 + \zeta_0\, \Psi_0]\,.
\ee
%%%%%
In a general dimension $n\ne4$ there are two obstructions to obtaining
a conserved charge.  Firstly, the term $\fft{n-4}{2}\, C_0^I\, \del_I\Psi_0$
is not a total derivative in general.  Note, however, that if
$C_0^I$ vanishes when $I$ lies in the direction(s) associated with 
latitude type coordinates on the sphere (i.e.~directions that are not
associated with  Killing vectors), then this term can be
rewritten as the total derivative $\fft{n-4}{2}\, D_I(C_0^I\,\Psi_0)$,
since the remaining, azimuthal, sphere coordinates are associated 
with Killing directions.  As we saw earlier, $C_0^I$ will indeed vanish
in the non-azimuthal directions in the case of the WAF metrics obtained
by conformal inversion of extremal black hole metrics, since $h_0^I=0$
in those directions in the black hole metrics.  This still leaves the
problem of the term $-\fft{(n-2)(n-4)}{4}\,\Psi_0 $ in (\ref{NPcon}).  This
term implies that there can be no Newman-Penrose charge for a massless
scalar obeying $\square\Psi=0$ in any dimension higher than $n=4$.

It is perhaps worth remarking that in the case of a WAF metric obtained by
conformal inversion of a static spherically symmetric extremal black hole
in $n\ge 5$ dimensions, the Ricci scalar goes like $1/r^2$ at large
distance and thus it would make a contribution in the NP charge calculation
at the
leading order in a large-$r$ expansion if one were to add an $R\,\Psi$ term to
the massless wave equation $\square\Psi=0$.  If the coefficient of this term
were chosen appropriately, it could be arranged to cancel the term 
$-\fft{(n-2)(n-4)}{4}\,\Psi_0 $ in (\ref{NPcon}), thus allowing the existence
of a conserved NP charge. (See eqn (\ref{5dR}) for the calculation of
the Ricci scalar term for the conformal inversion of the five-dimensional
extremal Reissner-Nordstr\"om metric.)  However, as may be readily checked,
the coefficient of $R\,\Psi$ that would be needed to achieve this cancellation
appears to have no other related significance.  In particular, it
is not equal to the coefficient that would be needed for the 
conformally invariant scalar operator. 

\section{Conclusions}

  In this paper, we have constructed the metrics in Gaussian null 
coordinates for certain classes
of extremal rotating black holes in supergravity theories, as expansions
in the radial coordinate at a sufficient order to be able to calculate
the conserved Aretakis charges on the horizon.  Specifically, we did
this for the extremal rotating black holes in four-dimensional STU
supergravity that carry four independent electric charges (with the
special case of the Kerr-Newman black hole when the four charges are
equal), and also for the
general extremal rotating 3-charge black holes in five-dimensional STU
supergravity.  We then obtained the explicit expressions for the
simplest of the Aretakis charges for a massless scalar field in each
case.

   We also investigated the possibility of relating the Aretakis
charge on the horizon of the extremal black hole to the Newman-Penrose charge
at $\scri^+$ in the metric obtained by performing an inversion of the
radial coordinate, after the extraction of an appropriate conformal
factor.  This relation was studied for four-dimensional spacetimes in
\cite{godgodpop}, where various examples of the mapping were obtained for
classes of static extremal black holes.  In the present paper we showed that
such a mapping becomes problematical for extremal rotating black holes,
because after conformal inversion the resulting weakly asymptotically 
flat metric has a Ricci scalar whose fall-off at large $r$ is 
sufficiently slow that one cannot treat the massless scalar equation 
$\square\psi=0$ as being equivalent to the conformally-invariant equation
$(\square-\ft16 R)\psi=0$ for the purpose of calculating the Newman-Penrose
charge.  This means that the ability to relate the solutions for the scalar 
field in the original extremal metric and in the conformally-inverted metric 
is lost in the case of rotating black holes, at least if we
consider just a simple $\rho\rightarrow 1/r$ inversion.  
In turn, one 
cannot by this means 
relate the Aretakis and Newman-Penrose charges for extremal rotating 
four-dimensional black holes.

As we then discussed, the situation becomes worse in dimensions $n>4$.  
The extremal black holes (static or rotating) still admit conserved 
Aretakis charges for a massless scalar field, but there are 
no Newman-Penrose charges for a massless scalar in any asymptotically flat
spacetime of dimension $n>4$.  Thus it appears that the mapping between
Aretakis and Newman-Penrose charges is exclusively a four-dimensional
phenomenon.

There remain a number of directions for further study.  Firstly, it
would be of interest to generalise the construction of the extremal rotating
four-dimensional STU black holes in Gaussian null coordinates to the
general case of eight charge parameters (independent electric and magnetic 
charges carried by each of the four gauge fields.  The solution for
the 8-charge rotating black holes is given in \cite{chowcomp1}.)  
It would also be of interest to study the analogous conserved 
Aretakis and Newman-Penrose charges for higher-spin fields in the
charged supergravity black hole backgrounds.  Examples would include
Maxwell fields, and also perturbations of the background metrics
themselves.

\section{Acknowledgements} We are grateful to Hadi Godazgar, Mahdi 
Godazgar, Carmen Li and James Lucietti for helpful discussions.
The work of M.C. is supported in part by the DOE (HEP) Award DE-SC0013528, 
the Fay R. and Eugene L. Langberg Endowed Chair (M.C.) and the Slovenian 
Research Agency (ARRS No. P1-0306).  
The work of C.N.P. is supported in part by DOE 
grant DE-FG02-13ER42020.

\end{document}